\documentclass{aastex631}

\usepackage{rotating}
\submitjournal{ApJ}

\shorttitle{The radial gradient of main sequence magnetic fields}
\shortauthors{M. Giarrusso}

\begin{document}
 
\title{Direct observation of the radial magnetic field gradient in HD\,58260\\
  from spectropolarimetry of NLTE lines in emission}

\author[0000-0002-4453-1597]{Marina Giarrusso} \email{marina.giarrusso@lns.infn.it}
\affiliation{INFN - Laboratori Nazionali del Sud, Via S. Sofia 62, I--95123 Catania, Italy}

\begin{abstract} 
  Because of the unquestionable presence of magnetic fields in stars, their role
  in the structure of stellar atmospheres has for a long time been a subject of
  speculation. In our contribution to this discussion we present spectropolarimetric
  evidence of the decrease of the radial component of the magnetic field with altitude
  in the atmosphere of HD\,58260, a B-type magnetic star on the main sequence.
  We show that the Stokes\,$V$ profiles of metal lines in emission of the outer
  atmosphere are evidence for a field three times weaker than from absorption lines from
  inner layers. The extra flow of energetic particles due to the magnetic-gradient
  pumping mechanism could be at the origin of the magnetospheres surrounding this class
  of stars and at the basis of the high energy phenomena observed. We also list a series
  of spectral lines useful for measuring the surface field of early-type stars.
\end{abstract} 


\section{Introduction}
The understanding of astrophysical plasmas has improved after the inclusion
of magnetic field effects in radiative transfer and polarization 
\citep{Landi2004}. The introduction of magnetic fields to the modeling of
atomic diffusion \citep{Alecian2019} also has provided precious insight into
the buildup of abundance anomalies. On the other hand, little has been done
to take into account the contribution of magnetic fields to the various
equilibria in stellar atmospheres. Only semi-quantitative approaches have
been proposed, with sometimes surprising conclusions. For example,
\cite{Hubbard1982} have claimed that the radii of magnetic Ap stars are
20\% larger than those of main sequence stars with the same temperature, a
result that would challenge our current view of the atmospheres of magnetic stars.

This paucity of efforts is a consequence of the lack of observational evidence
of a radial gradient in the field strength that is necessary to give rise to a
magnetic pressure. \cite{Wolff1978} made the first-ever attempt to derive the
magnetic gradient in stellar atmospheres by measuring the longitudinal component
of the field from lines formed from optical depth $\log\tau_{\rm 500\,nm}$ = 0.42
to 0.03. It was found that variations, if present at all, are not larger than 20\%.
In similar studies, \cite{Nesvacil2004, Romanyuk2004, Kudryavtsev2009} have
reported magnetic fields $5-25$\% stronger in the inner layers (where spectral
lines shortward of the Balmer jump are formed) with respect to the outermost
layers (where spectral lines longwards of the Balmer jump come from). The
investigation by \cite{Wolff1978} was hampered by the modest depth of the visible stellar disk
$T_{\rm eff}\sim8000$\,K atmospheres examined ($\log\tau_{\rm 500\,nm}$ changes
from 1 to $-3$ over just 1600\,km).
 
Early B-type stars are more suitable objects
to look for a field gradient; their outer layers turn ``visible" when metal
spectral lines are in emission. \cite{Sigut2000} reported for the first time the presence of metal lines in emission in the visible spectrum of B-type stars. These authors suggested that in this class of stars emission lines are produced by Non-Local Thermodynamic Equilibrium (NLTE) effects because of source functions rising with the atmosphere height without the necessity of a temperature inversion (chromosphere). 
A conclusion theoretically supported by \cite{Alexeeva2016, Alexeeva2018, Alexeeva2019, Mashonkina2020, Mashonkinaetal2020}.
From an analysis of line profiles of a sample of early B-type stars, \cite{Sadakane2017} concluded that emission and absorption lines
are equally broadened by the stellar rotation, supplying an observational proof that emission lines are from the outer stellar atmosphere and not from a circumstellar environment.

The present paper reports the results of a search for a radial gradient in the
magnetic field of the early B-type star HD\,58260, based on high-resolution Stokes
$I$ and $V$ profiles of spectral lines in absorption and in emission, sampling at
least 6\,dex in optical depth. 
This magnetic \citep{Bohlender1987} helium-rich \citep{Garrison1977} star has been selected because of its very low projected rotational velocity ($v_{\rm eq}\sin\,i\,= 3$ km\,s$^{-1}$) and absence of evidence for either magnetic or light variability \citep{Shultz2018, Shultz2019}.
All these properties simplify the procedure of measuring the magnetic field from spectral lines and the interpretation of results. 

\begin{deluxetable*}{lcccccccccc}\label{twocomp}
\tablecaption{Atomic transitions -- resulting in a Zeeman doublet with two $\pi$ components in
    coincidence with two $\sigma$ components -- suitable for measuring the surface magnetic
    field of B-type stars. Line-by-line $B_{\rm s}$ measurements are from a double Gaussian function fit to the two components
    (see Figure\,\ref{SII5473}).}
\tablewidth{0pt}
\tablehead{
\colhead{Ion}  &  \colhead{\AA}   & \colhead{Configuration} & \colhead{Term} & \colhead{J} & \colhead{$g_L$} & \colhead{Configuration} & \colhead{Term} &  \colhead{J} & \colhead{$g_L$} &  \colhead{$B_{\rm s}$ (\rm G)}  \\
}
\startdata
C {\sc ii}  &6787.207    & $2s^2 2p^2\,(^3P^o)\,3s$ & $^4P^o$ & 1/2 & 2.67 & $2s^2 2p^2\,(^3P^o)\,3p$ & $^4D $ &1/2 & 0.00 & 4200$\pm$140 \\
Si {\sc ii} &6699.401 & $3s\,\, 3s^3\,(^3P^o)\,4s$ & $^4P^o$ & 1/2 & 2.67 & $3s\, 3p^2\,(^3P^o)\,4p$ & $^4D$   &1/2 &0.00 & 4320$\pm$130 \\
S {\sc ii} &4278.516 & $3s^2 3p^2\,(^3P)\,4p$ & $^4P^o$ & 1/2 & 2.67 & $3s^2 3p^2\,(^3P)\,4d$ & $^4D^o$   &1/2 &0.00 & 4700$\pm$160 \\
S {\sc ii} &4456.382 & $3s\,\, 3s^2\,(^3P)\,4p$ & $^4D^o$ & 1/2 & 0.00 & $3s^2 3p^2\,(^3P)\,5s$ & $^4P$   &1/2 &2.67 & 4400$\pm$270 \\
S {\sc ii} &5473.614 & $3s^2 3p^2\,(^3P)\,4s$     & $^4P$    & 1/2 & 2.67 & $3s^23p^2\,(^3P)\,4p$    & $^4D^o$   &1/2 & 0.00 & 4440$\pm$40\,\,\\\hline\enddata
\end{deluxetable*}

\section{The magnetic Field of HD\,58260}
Hereafter, definitions and measuring procedures of stellar magnetic fields are as in
\cite{Leone2017}. We just recall: 1) the effective -- or longitudinal -- magnetic field
($B_{\rm eff}$) is the average across the visible stellar disk of the longitudinal
component of the field and 2) the surface magnetic field ($B_{\rm s}$) is the average
across the visible stellar disk of the field modulus.

\subsection{Effective Magnetic Field and Variability Period}

\cite{Shultz2018} measured the effective magnetic field of HD\,58260 from 
ten high-resolution spectra in circularly polarized light collected with ESPaDOnS
\citep[${\rm R}=\lambda/\Delta\lambda=68\,000$,][]{Donati2006, Wade2016}, between MJD = 56623 and 56678. 
With the inclusion of literature data, these authors conclude that this helium-rich star presents a $B_{\rm eff}$ = 1820$\pm$30\,G
over a time-scale of $\sim\,$35 years, a lack of variability corroborated by the {\it Hipparcos} photometry.
Within the framework of the Oblique Rotator Model \citep{Babcock1949}, 
the constant magnetic field and the shape of the Stokes $V$ profiles are due
to the coincidence of the stellar rotational axis with the line-of-sight (LoS) \citep{Shultz2019}. 

From a further HARPS \citep[${\rm R}=115\,000$,][]{Mayor2003} spectrum in circularly polarized light 
(MJD = 57091.1462), \cite{Jarvinen2018} measured $B_{\rm eff} = 1537\pm37$\,G 
and classified HD\,58260 as a magnetic variable. Even though this is possible, we note that
both \cite{Shultz2018} and \cite{Jarvinen2018} measured $B_{\rm eff}$ from the ``average''
least-squares deconvolution profile \citep{Donati1997} under the assumption that Stokes $V$ is the first
derivative of Stokes $I$, and neglecting the fact that the line profiles of HD\,58260
are shaped not only by the magnetic field but also by stellar rotation and instrumental
broadening. \cite{Scalia2017} and \cite{Ramirez2020} have performed numerical simulations
to check the importance of these extra broadening mechanisms when measuring stellar
magnetic fields. Here it is appropriate to quote one of \cite{Scalia2017} results: a
$B_{\rm eff} = 3175$\,G field would appear as 2495\,G at R = 65\,000 and an instrumental
profile with FWHM = 2.5 pixels, equal to 1697\,G with R = 115\,000 and FWHM = 4 pixels. 
In addition, \cite{Shultz2018} and \cite{Jarvinen2018} could give discrepant results because
of different line lists differently sampling the non homogeneous distribution of elements on the surface of HD\,58260.
It is thus possible that the differences between ESPaDOnS and HARPS results are not
conclusive and so the variable nature of HD\,58260 remains to be confirmed.

Hereafter, we refer to the average Stokes $I$ and $V$ spectra of HD\,58260 computed from the observations presented by
\cite{Shultz2018}. The Signal-to-Noise ratio is of the order of 1000 in the Stokes $V$ continuum.
\begin{figure}
\includegraphics[width=8.5cm]{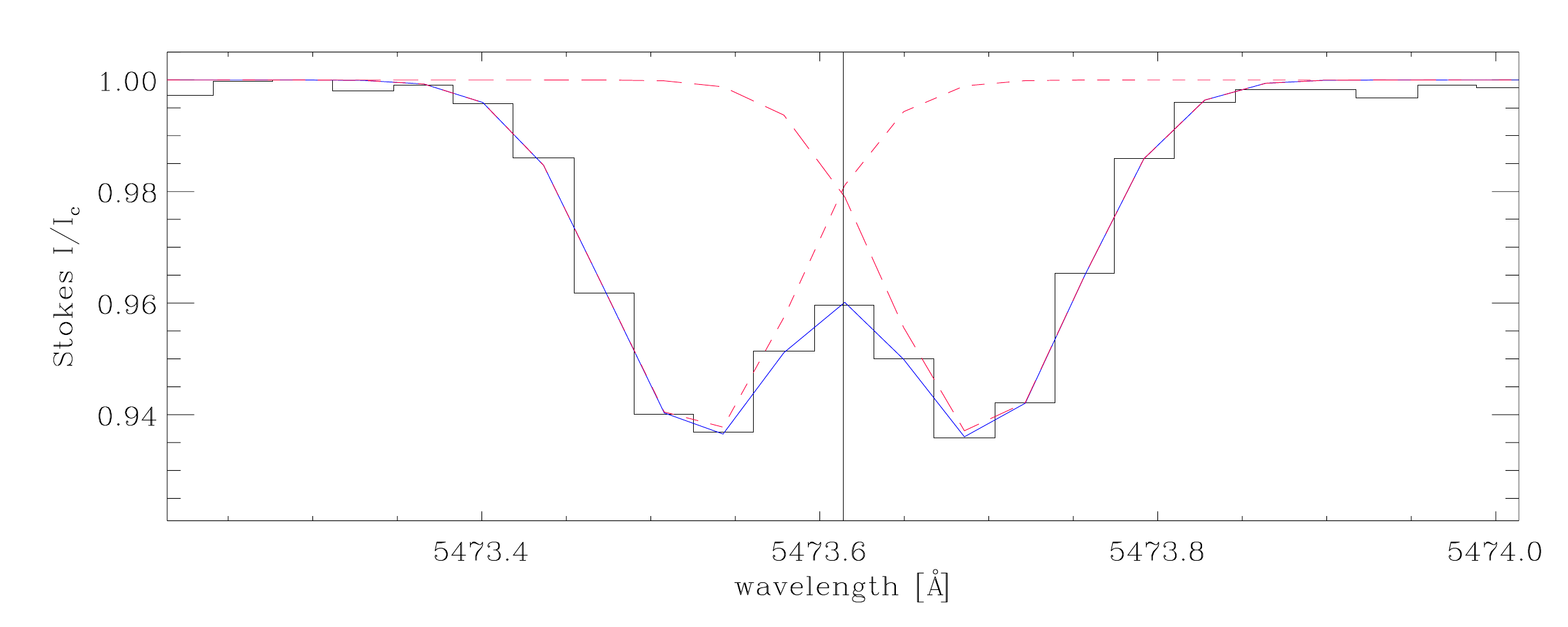}
\caption{Histogram of the observed S{\sc II} 5473.614\,\AA\, line. This splits in just two
  Zeeman subcomponents fitted with a double Gaussian function (blue-solid). In this case the distance
  between the two (red-dashed) components corresponds to $B_{\rm s} = 4440\pm40$\,G, as listed in Table\,\ref{twocomp}.}
\label{SII5473}
\end{figure}

\subsection{The Surface Magnetic Field}
Following \cite{Mathys1990}, there is agreement that $B_{\rm s}$ is best measured using 
Fe{\sc ii}\,6149.258\,\AA, a spectral line representing a simple doublet arising
from coinciding $\pi$ and $\sigma$ subcomponents. Since this line is absent from the
spectra of early type stars, we have mined from the NIST five spectral lines 
present in the spectrum of HD\,58260 that exhibit equally lucky quantum numbers to result in
Zeeman doublets (Table\,\ref{twocomp}). Fitting the observed doublets
with double Gaussian functions (as shown in Figure\,\ref{SII5473} for S{\sc II} 5473.614\,\AA\, line) we have obtained the $B_{\rm s}$ values listed in Table\,\ref{twocomp}. The weighted average gives a surface field of HD\,58260 equal to $B_{\rm s} = 4430\pm40$\,G.

\begin{figure}
\gridline{\fig{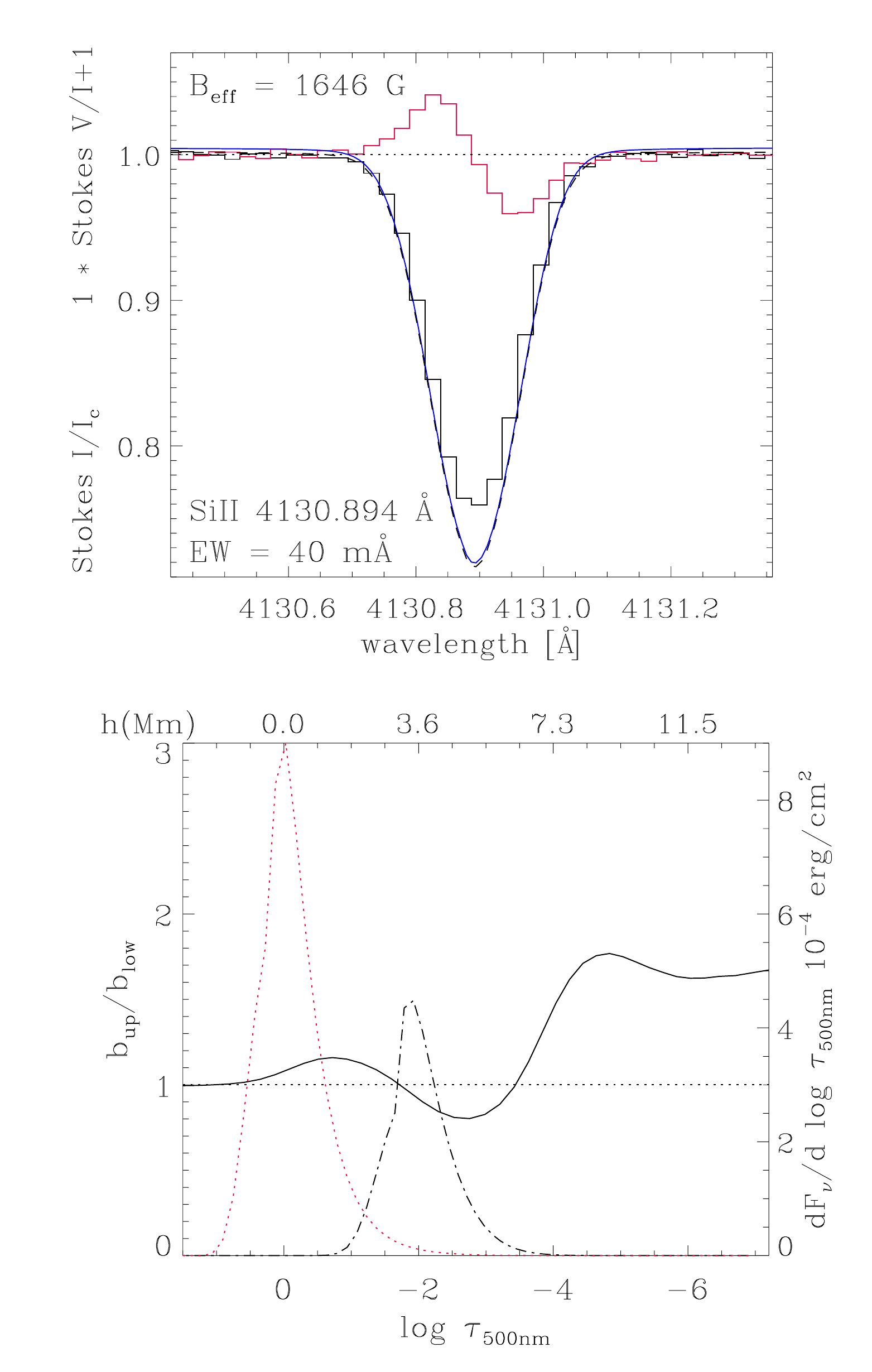}{0.25\textwidth}{(a)}
          \fig{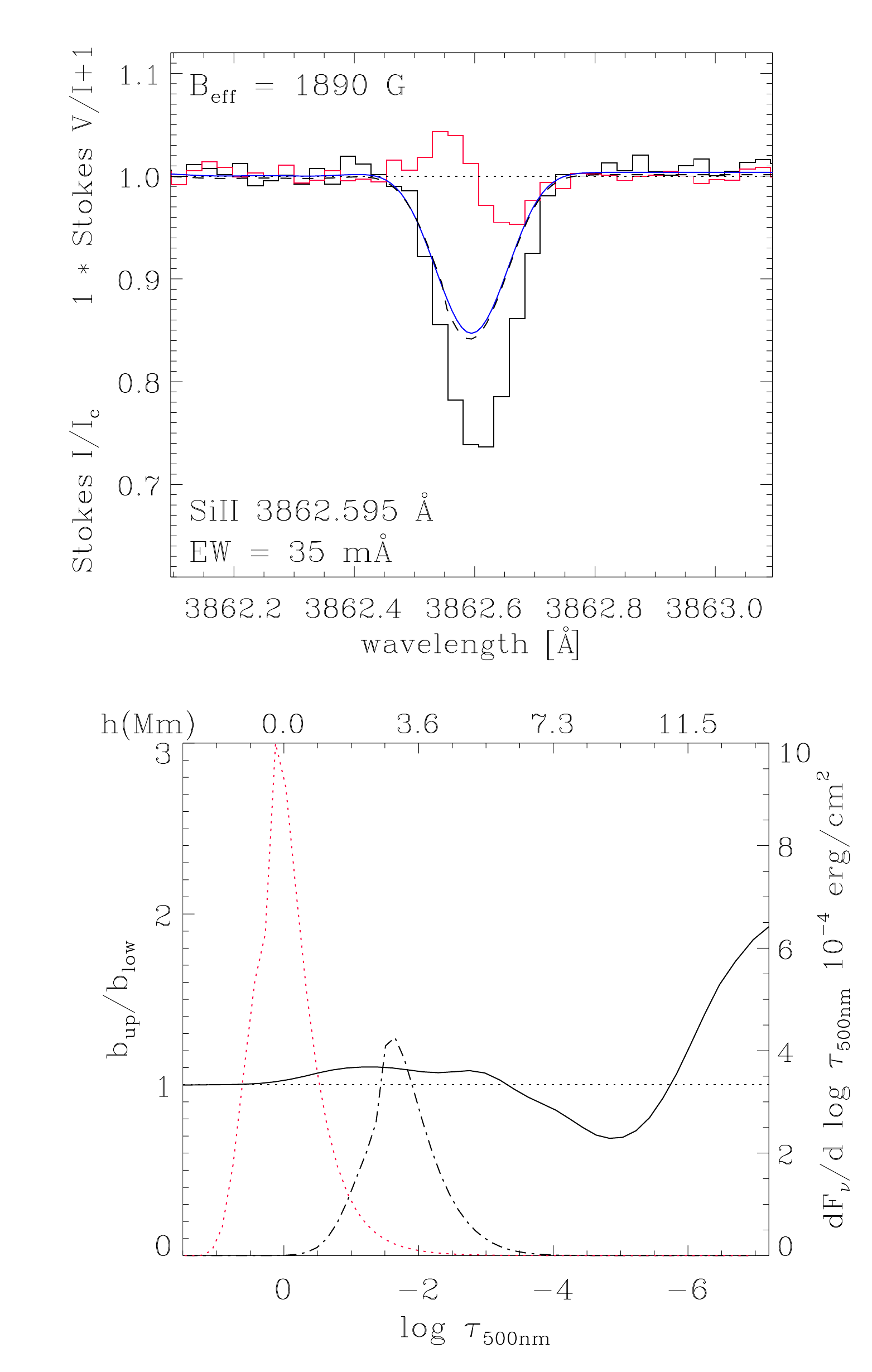}{0.25\textwidth}{(b)}
          \fig{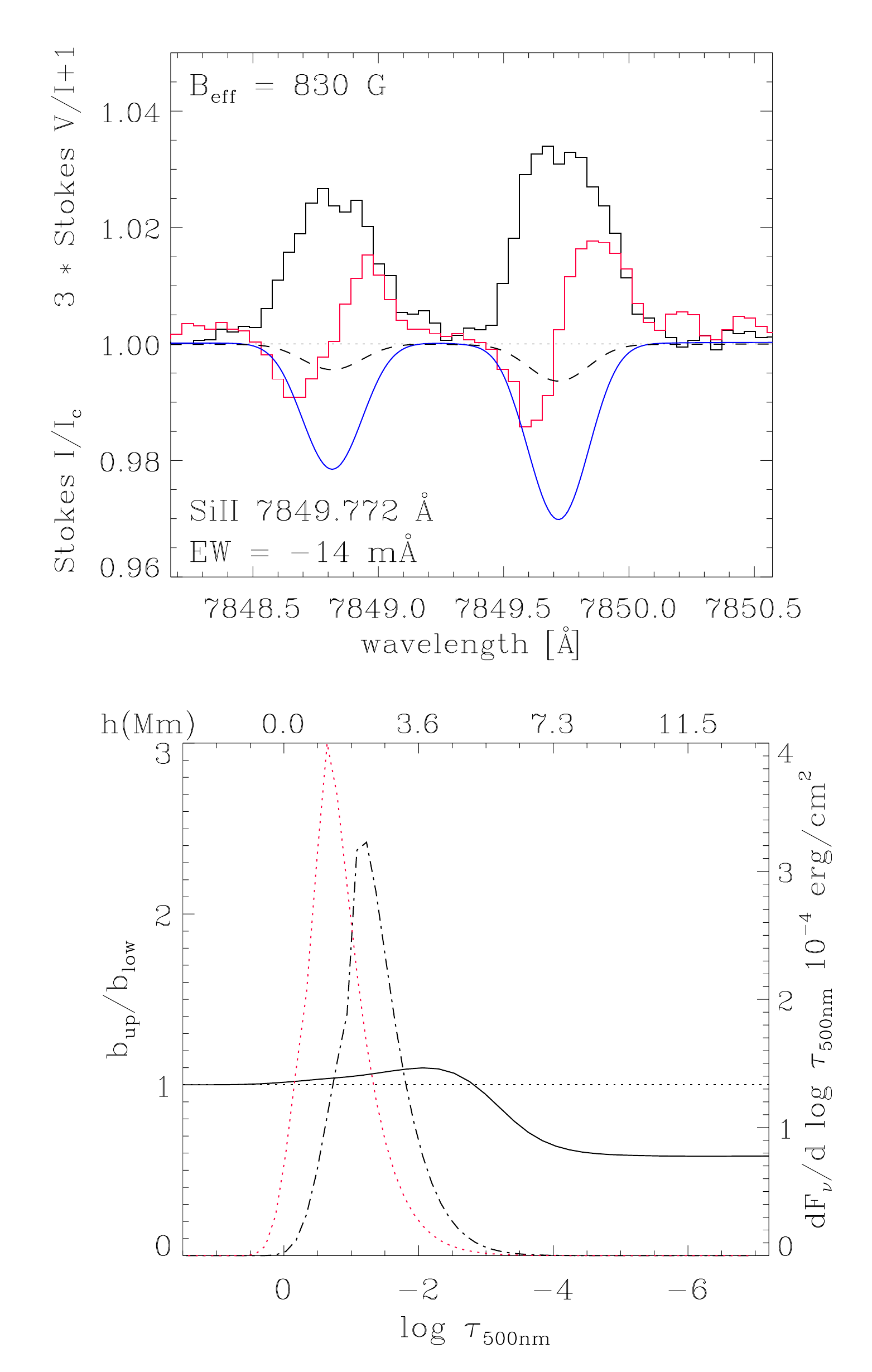}{0.25\textwidth}{(c)}
          \fig{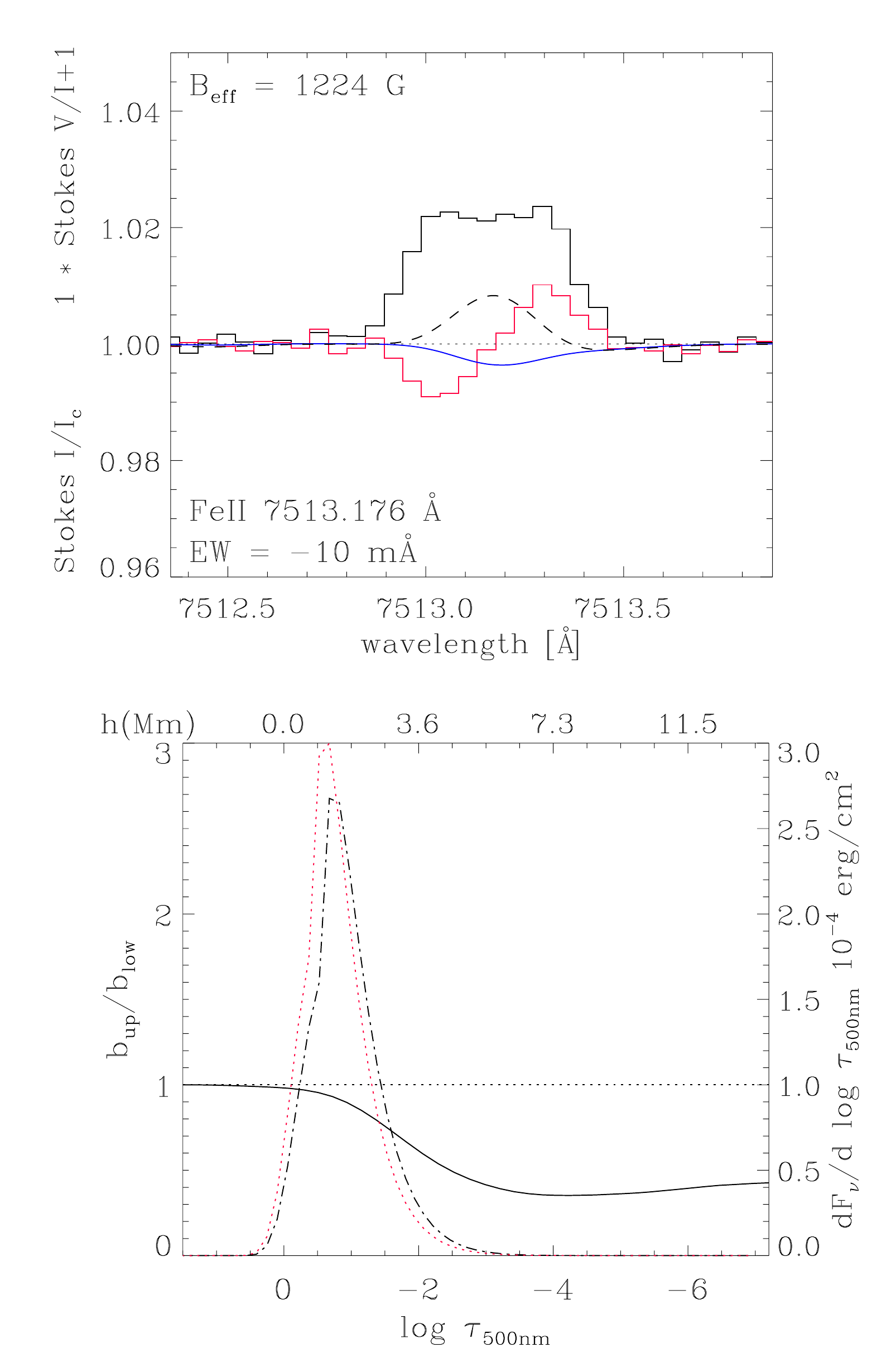}{0.25\textwidth}{(d)}
          }
\caption{Top: examples of Stokes $I/I_{\rm c}$ and $V/I$ observed profiles (offset by unity), displayed as black and red histograms respectively, of lines in absorption (a, b) and emission (c, d). LTE (full) and NLTE (dashed) synthetic profiles are also reported. Bottom: \cite{Gray1992} contribution function for both the line core (black dash-dotted) and the line continuum (red dotted), and ratio b$_{\rm up}$/b$_{\rm low}$ between the departure coefficients of the involved levels (black solid), as a function of optical depth and geometrical height h. See Section\,\ref{lbl} for details.}
\label{righe}
\end{figure}

\subsection{Magnetic Field Geometry}\label{submfg}
As to a dipole with polar strength $B_{\rm p}$, \cite{Preston1969} has shown:\\ 
1) for a LoS coincident with the dipole axis
\begin{equation}\label{equ1}
B_{\rm s} = 0.68 \frac{3.42 - u}{3 - u} B_{\rm p} \hspace{0.3cm} {\rm and} \hspace{0.3cm} B_{\rm eff} = \frac{1}{20}\frac{15 + u}{3 - u} B_{\rm p}
\end{equation}
\\
2) when the LoS lies in the magnetic equatorial plane 
\begin{equation}
B_{\rm s} = 0.68 \frac{2.76 - u}{3 - u} B_{\rm p} \hspace{0.3cm} {\rm and} \hspace{0.3cm} B_{\rm eff} = 0
\end{equation}
Here, $u$ is the linear limb-darkening coefficient.

If the magnetic field of HD\,58260 is dipolar and the LoS coincident with the
rotation axis \citep{Shultz2018}, the ratio $B_{\rm eff}/B_{\rm s} = 1860/4430 \sim  0.42\pm0.06$
-- which is close to the highest possible value of 0.37 \citep[$u = 0.4$,][]{Neiner2003} -- also implying the
coincidence of dipole and rotation axes and a polar field $B_{\rm p}\sim6000$\,G 
(consistent with the geometry and surface field strength determined by \cite{Shultz2019}). 
See Section\,\ref{lbl} for the adopted $B_{\rm eff}$ value.

\section{$B_{\rm eff}$ from emission lines}\label{lbl}
We selected the magnetic early B-type star HD\,58260 for the presence of visible metal
lines in emission, unquestionably with circularly polarized profiles (Figure\,\ref{righe}).
Keep in mind that Stokes\,$V$ profiles of emission lines are reversed in sign with respect
to absorption lines. 
\cite{Wade2012} already reported on the presence of not null Stokes $V$
profiles in emission in the spectrum of the O7f?cp star NGC 1624-2. These authors
also found a Stokes $V$ change in sign between absorption and weak-emission O{\sc iii} lines and propose to locate
the formation region of these emission lines in the magnetosphere.   

We ran the {\sc tlusty205} and {\sc synspec5.1} codes \citep{tlusty2007} to model -- under
the assumption of both LTE and NLTE -- the spectral lines of HD\,58260 for T$_{\rm eff}$ = 19000\,K,
$\log g = 4.0$ \citep{Sigut2000} and solar abundances. No attempt has been made to
determine the stellar parameters by matching the line strengths. Computations were aimed
at line identification and evaluation of departure coefficients. Figure\,\ref{righe} shows
examples of observed spectral lines both in absorption and emission together with:
{\it i)} the theoretical spectra computed in LTE and in NLTE, {\it ii)} the \cite{Gray1992} contribution
functions (which give the relative contribution of the different atmospheric layers)
 in the line center and in the corresponding continuum -- computed in LTE
-- plotted vs. the optical depth at $\tau_{\rm 500\,nm}$ (and the geometrical height), and
{\it iii)} the ratio b$_{\rm up}$/b$_{\rm low}$ between the departure coefficients of the levels involved in the transition, 
where b = n$_{\rm NLTE}$/n$_{\rm LTE}$ (being n$_{\rm NLTE}$ and n$_{\rm LTE}$ the actual population and the LTE population of a level, respectively). 
With a value $< 1$, lines in emission have an NLTE origin in the outer atmospheric layers of HD\,58260.
 
We measured $B_{\rm eff}$ line by line both from absorption and emission profiles, using
the first order momentum of Stokes $V$ \citep{Mathys1994}. From 16 lines in emission
(Table\,\ref{em_line}) we obtained  an average value of $B_{\rm eff}$ = 675$\pm$260\,G.
The average from 84 absorption lines is $B_{\rm eff} = 1860\pm230$\,G. 
Estimated from equation\,15 by \cite{Mathys1994}, errors are $\propto EW^{-1}$ and dominated here by the
very small equivalent widths of lines selected to sample the optical depth as widely
as possible. To test the hypothesis that our sample of $B_{\rm eff}$ measurements from
emission lines was drawn at random of the same population of $B_{\rm eff}$ measurements from
absorption lines, the Student $t-$statistics has been applied. The significance of 2.7$\times 10^{-33}$
indicates that the $B_{\rm eff}$ measurements from emission and absorption lines have different means. 
Also \cite{Wade2012} found $B_{\rm eff}$ from weak-emission lines about a factor 2 smaller than from absorption lines.

\begin{figure}
\plotone{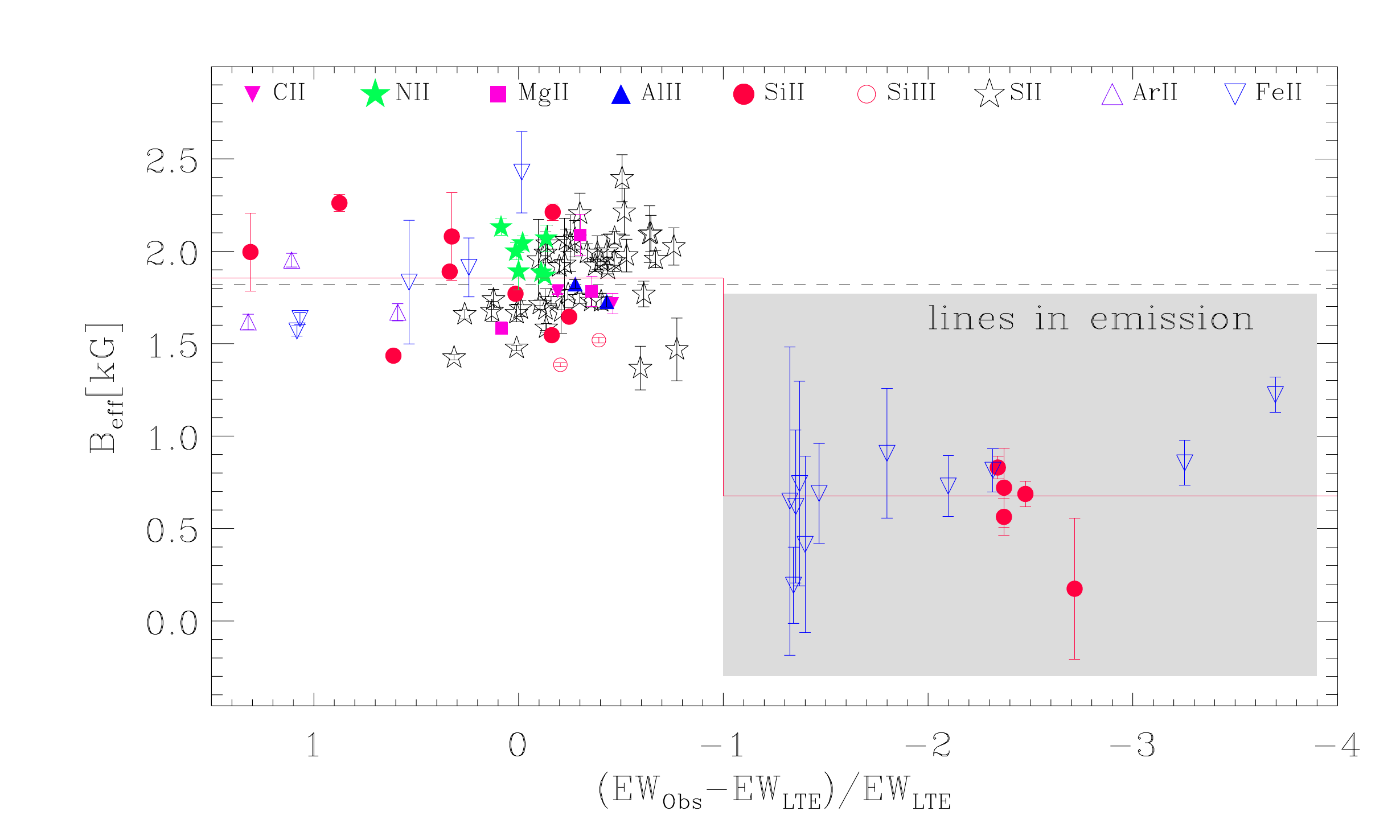}
\caption{Line by line measurements of $B_{\rm eff}$ from nine different species
  vs. the relative differences between measured (EW$_{\rm Obs}$) and theoretical
  LTE (EW$_{\rm LTE}$) equivalent widths. The smaller the abscissa value, the further out the line formation region. 
  The filled region contains lines in emission. The dashed line represents the field as measured by \cite{Shultz2018}.
  The  broken line goes through the average field values for absorption and emission spectral lines.}
\label{gradiente}
\end{figure}

Locating the region of the formation of a line in emission depends on the adopted
atmospheric model and the atomic parameters. Figure\,\ref{righe}(c) shows the present
limit of NLTE computations in the case of the two silicon lines in emission at
7849\,\AA. Theoretically, the population levels are significantly inverted locating the formation regions of these two lines at 
$\log\tau_{\rm 500 nm} < -4$ (bottom panel),
but with {\sc synspec5.1} these lines are still in absorption (top panel). Even worse, not
every line can at present be treated under the NLTE approximation. For a
quantitative representation of NLTE effects on the 100 examined spectral lines with clearly
observable Stokes $V$ profiles we establish the relative difference between the
observed equivalent width (EW$_{\rm Obs}$) and the expected value in the LTE
approximation (EW$_{\rm LTE}$):  $r = $(EW$_{\rm Obs} - $EW$_{\rm LTE}$)/EW$_{\rm LTE}$. 
With LTE lines always in absorption, the zero value of the $r$-scale depends only on the observed lines and the scale can only be shrinked or expanded by assuming a different temperature or surface gravity of the LTE atmosphere.
The idea behind this: the smaller $r$, the further out the line formation region.
Figure\,\ref{gradiente} visualizes the dichotomy between the $B_{\rm eff}$ values from the
84 absorption ($r > -1$) lines and the 16 emission ($r < -1$) lines respectively.

For lines in emission, inverted populations of the atomic levels at $\log \tau_{\rm 500\,nm}<-4$ is
corresponding to an altitude of tens of Mm above the layers with unit optical depth, where absorption lines
are mainly formed (see Figure\,\ref{righe} showing their contribution functions). The scale of the vertical
gradient of the magnetic field is $\sim$\,0.1\,G\,km$^{-1}$.
This is a value much larger than what results for a magnetic dipole, for which it would be necessary to go to a distance of
0.4\,$R_*(\sim 0.4\,\times 3 R_\sun \approx 840$\,Mm) from the surface of HD\,58260
to experience a decrease of the field to 600\,G from the surface value of 1800\,G. 
An unrealistic extended atmosphere for the current view of magnetic early-type stars whose
radii are not too different than for main-sequence stars with equal temperature \citep{Shultz2019b}. 

\begin{figure}
\plotone{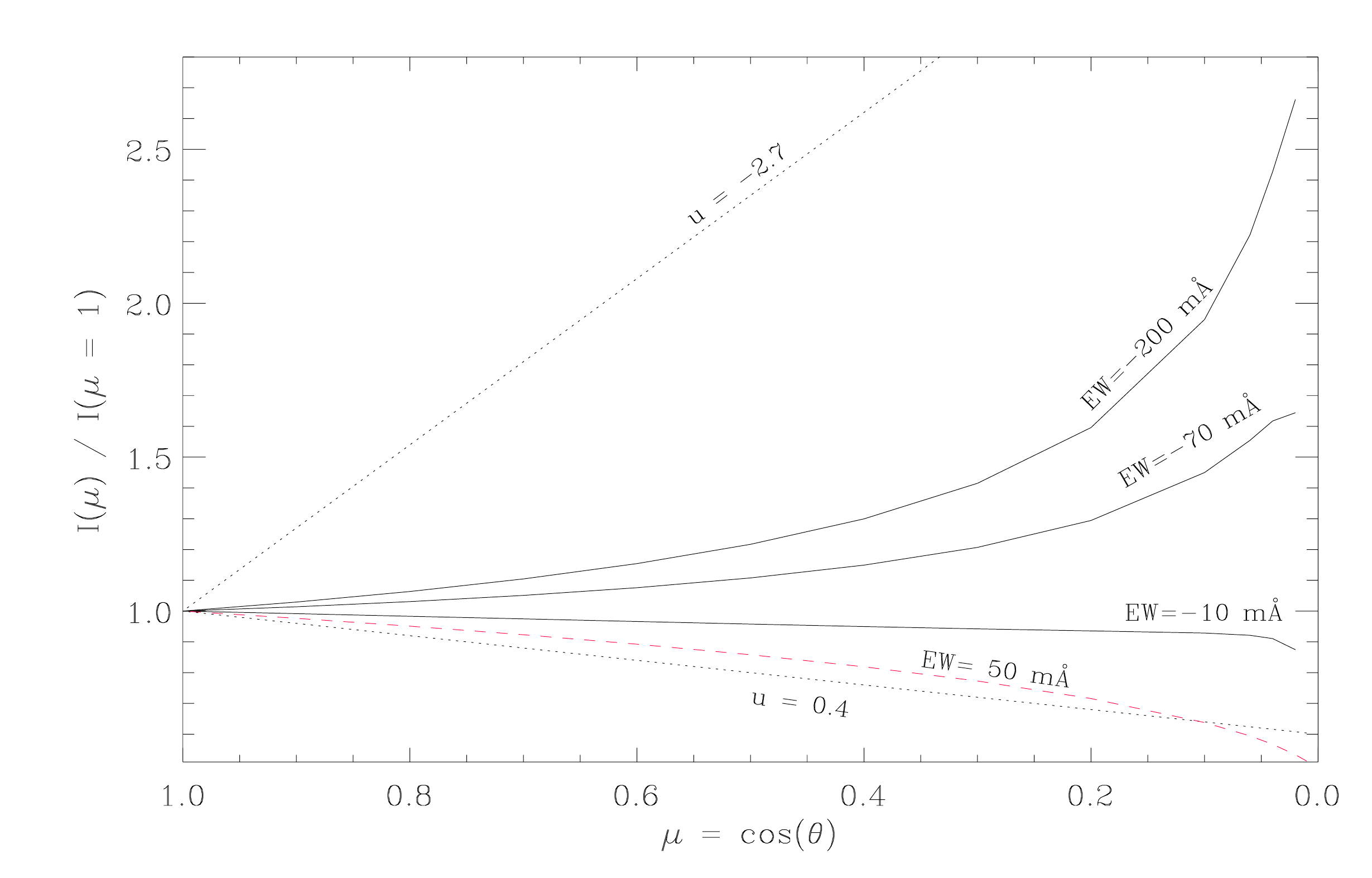}
\caption{{\sc synspec5.1} intensity, across the stellar disk, of the Fe{\sc ii} 7513.176\,\AA\, spectral line in emission (negative Equivalent Widths, EW) for different
abundance values. $\theta$ is the angle between LoS and the direction of the emerging flux. For comparison, the intensity variation of the LTE line in absorption (EW = 50 m\AA; red-dashed) is reported together with (dotted) linear ($u=0.4$) limb-darkening and ($u=-2.7$) limb-brightnening.}
\label{limb}
\end{figure}

Before drawing specific conclusions about the large difference between $B_{\rm eff}$ values from spectral lines in absorption (1860$\pm$230\,G) and in emission (675$\pm$260\,G), the importance of limb-darkening (see Section\,\ref{submfg}) has to be understood. 
From previous \cite{Preston1969} relations (Eqs.\,\ref{equ1}), a {\it limb-brightening} for $u \sim -2.7$ could analytically reconcile this difference. A possibility ruled out by {\sc synspec5.1} intensities of spectral lines: limb-brightening (Figure\,\ref{limb}) due to NLTE effects is not so large  to justify the measured differences in the longitudinal field from lines in absorption and emission.
Figure\,\ref{limb} also shows an almost constant brightness across the visible disk for lines in emission of a
few $m$\AA\, eventually resulting in $B_{\rm eff} \sim 1550$ G. In conclusion, in the framework of present NLTE modeling of early B-type stars, the longitudinal component of the photospheric magnetic field of HD\,58260 is decreasing with the stellar radius.

\begin{deluxetable*}{lcrcrc}\label{em_line}
\tablecaption{Spectral lines in emission from which it was possible to measure the
    effective longitudinal field $B_{\rm eff}$ (in Gauss). Measured equivalent widths 
    EW (in $m\rm \AA$) and adopted effective Land\'e factors $g_{\rm eff}$ are also listed, 
    together with the Chi-Square Probability Function $p$ of the single-line measurements from the mean spectrum as in \citet{Donati1992}: 
    magnetic field detection (D) for $p>0.99$, marginal detection (MD) for $0.95<p<0.99$, no detection (ND) for $p<0.95$.}
\tablewidth{0pt}
\tablehead{
\colhead{Ion}  &  \colhead{\AA}  & \colhead{EW} & \colhead{$g_{\rm eff}$} & \colhead{$B_{\rm eff}$} & \colhead{$p$} \\  
}
\startdata
Si II  &  6239.665 & $-$13.4  &  0.93  &  685$\pm$140  &  D \\ 
Si II  &  6818.414 & $-$2.2   &  0.83  &  175$\pm$760  &  D \\ 
Si II  &  6829.799 & $-$3.4   &  0.83  &  720$\pm$430  &  D  \\ 
Si II  &  7848.816 & $-$10.6  &  0.90  &  560$\pm$190  & D  \\ 
Si II  &  7849.722 & $-$14.4  &  1.06  &  830$\pm$125  &  D \\ 
Fe II &  4948.791 & $-$0.6   &  1.14  &  650$\pm$870  &  ND \\
Fe II &  5144.352 & $-$1.2   &  0.89  &  745$\pm$700  &  ND \\
Fe II &  5247.956 & $-$1.8   &  0.73  &  620$\pm$825  &  D \\
Fe II &  5251.226 & $-$2.2   &  1.48  &  195$\pm$415  &  ND \\
Fe II &  5429.987 & $-$1.7   &  1.06  &  690$\pm$540  &  ND \\
Fe II &  5493.831 & $-$1.3   &  1.06  &  905$\pm$700  &  D \\
Fe II &  5780.128 & $-$1.4   &  1.00  &  415$\pm$955  &  D \\
Fe II &  7513.176 & $-$10.4  &  1.30  & 1225$\pm$200  &  D \\
Fe II &  8490.102 & $-$6.7   &  1.03  &  855$\pm$245  &  D \\
Fe II &  8499.617 & $-$4.3   &   0.89 &  815$\pm$235  &  MD \\
Fe II &  8722.461 & $-$4.5   &   1.10 &  730$\pm$330  &  D \\
\enddata
\end{deluxetable*}

\section{Discussion and conclusions}
We have measured the surface magnetic field of HD\,58260  from newly
identified spectral lines that split into a Zeeman doublet
with the $\sigma$ components in coincidence with the $\pi$ components (see
Table\,\ref{twocomp}). These doublets are in general suitable for measuring
the surface field of early type stars, as the Fe{\sc ii} 6149.258\,\AA\, one is for cool magnetic stars.
Obtained from spectral lines in
absorption, the measured ratio $B_{\rm eff}/B_{\rm s}$ favors a coincidence
of line-of-sight, rotation axis and dipole axis; the polar field value is of the order of 6000\,G.

The metal emission lines of HD\,58260 constitute extreme evidence for the
breakdown of the LTE approximation in early type magnetic stars, but absorption
lines are affected by NLTE too. Systematic differences in the optical depth between observed and LTE equivalent
widths in magnetic stars, as a class, are commonly interpreted as due to a vertical
stratification of elements \citep{Catanzaro2016, Catanzaro2020}. 
Nevertheless, for these stars the attempt to reproduce observed metal emission lines by taking into account the NLTE
effect works in the direction to smooth abundance gradients \citep{Alexeeva2016, Alexeeva2019, Mashonkina2020, Mashonkinaetal2020}.
Remaining differences between observed and theoretical lines could be a matter of present limits of NLTE modeling 
that needs to be improved for treating the upper atmospheric regions, 
or the consequence of level populations distorted by the far-UV \citep{Wahlgren2004} and/or X-ray \citep{Mitskevich1992}
radiation from the magnetosphere wrapping this class of stars \citep{Shore1990}. HD\,58260
appears an example of the necessity for a simultaneous and integrated study of atmosphere and magnetosphere.

HD\,58260 presents emission lines with non-null Stokes\,$V$ profiles. Determined
via the first-order momentum of these profiles \citep{Mathys1994}, the measured
longitudinal magnetic field from these lines ($\sim$\,675\,G) is three times weaker
than from lines in absorption ($\sim$\,1860\,G). Following \cite{Sadakane2017}, we conclude that emission lines
originate from the outermost layers and that the magnetic
field decreases with altitude in the atmosphere more fastly than expected for a dipolar field.
In the current state of NLTE modeling, the radial magnetic field gradient in HD\,58260 is
real and it cannot be ascribed to an incorrect limb-brightening.

No conclusion concerning the decrease in field strength with altitude is possible
unless new observations of the Stokes $Q$ and $U$ profiles provide us with the
transverse component of the magnetic field \citep{Leone2017}. 
At this stage, it is not possible to foresee the consequence of a more compact magnetosphere 
on the synchrotron and auroral radio emission \citep{Leto2021} from this class of stars.

The estimated scale of the vertical gradient of the magnetic field of HD\,58260 is $\sim$\,0.1\,G\,km$^{-1}$.
The Sun is the only other astrophysical place where the vertical gradient of
the magnetic field has been observed: 1\,-\,4\,G\,km$^{-1}$ in solar spots
\citep{Joshi2017} and 10$^{-4}$\,G\,km$^{-1}$ in the corona \citep{Gelfreikh1997}. 
Indeed the resemblance between HD\,58260 and the Sun could go further.
\cite{Tan2020} have shown that the magnetic-gradient pumping mechanism can
power energetic particle up-flows that are responsible for solar eruptions.
These authors have also shown that the condition for magnetic-gradient pumping is the possibility for a
charged particle to make at least a complete circle around magnetic lines before a collision with other particles. In practise, 
the gyro-frequency has to be larger than collision rate: $\nu_g > \nu_c$.
The very presence of lines in emission is proof of an electron collision rate smaller than Einstein
spontaneous-emission probability ($A_{ji}$, for transition $j \rightarrow i$), so that magnetic-gradient pumping
is certainly at work in the case $\nu_g > A_{ji} > \nu_c$, with $A_{ji} \approx 10^7$\,s$^{-1}$ for lines
listed in Table\,\ref{em_line}. As to electrons, the non-relativistic gyro-frequency $\nu^e_g$ (= 2.8\,MHz\,G$^{-1}$) is equal to 16.8 GHz at the
magnetic pole, that is $\nu^e_g \approx 10^3 A_{ji}$. In a completely ionized plasma,
proton collision rate is 60 times smaller than electron one and this leads to $\nu^p_g \approx 17 A_{ji}$.
It appears that not only protons but also many light ions can be subject to the magnetic-gradient pumping mechanism in an early B-type
star like HD\,58260. The less demanding case $A_{ji} > \nu_g > \nu_c$ is more difficult to be evaluated.

HD58260 presents UV spectral lines in emission \citep{Shore1990} that are an unquestionable evidence of a
high-temperature plasma wrapping the star. However, as already known from \cite{Pedersen1979}, we found no
H$_\alpha$ emission in the here analyzed ESPaDOnS spectra (according to \cite{Shultz2019}), neither X-ray emission has been reported by \cite{Petit2013}
or radio emission by \cite{Linsky1992, Kurapati2017}. A behaviour different than other similar stars that can present some or even all these phenomena,
like $\sigma$\,Ori\,E showing X-ray emission \citep{Sanz2004},
UV lines \citep{Shore1990}, H$_\alpha$ emission \citep{Pedersen1979} and radio emission \citep{Cassinelli1985}.
 If such a pumping mechanism is indeed at work in magnetic stars with high-energy
emission phenomena (e.g. X-ray, UV spectral lines of highly ionized species,
H$_{\alpha}$ \citep{Leone2010} and radio), many unsolved problem can be addressed.
The extra flow of energetic particles and a magnetic gradient different than for a dipole one
could explain why magnetospheres are not ubiquitous. Magnetospheres would be the result of the interplay of
1) accelerated charges able, at short stellar distances, to ionize at least up to Si III (33.5 eV) and C III (47.9 eV), and at large distances to contribute to the filling of the H$_{\alpha}$ ($<$\,13.6 eV) emitting torus; and 2) field gradient, changing the location of the H$_{\alpha}$ emitting torus or even preventing its existence, as well the position of the centrifugal breakout \citep{Havnes1984}. Pumping could also play a role
in the stellar wind mass-loss rates necessary to explain the observed
radio emission from rotating magnetospheres. These $10^{-10}$\,-\,$10^{-9}M_{\odot}yr^{-1}$ rates \cite[see][for
a complete list of references]{Leto2020b} are much larger than predicted for main sequence
B-type stars \citep{Kritcka2014} or observed ($<10^{-12} M_{\odot}yr^{-1}$) in
the UV spectrum of the radio source CU\,Vir  \citep{Kritcka2019}.

\section*{Acknowledgments} We acknowledge financial contribution from the agreement ASI-INAF n.2018-16-HH.0. 

\bibliography{HD58260_V3}{}
\bibliographystyle{aasjournal}

\end{document}